\documentclass[11pt]{article}
\usepackage{amssymb}
\usepackage{epsfig}
\newlength{\bredde}
\def\slash#1{\settowidth{\bredde}{$#1$}\ifmmode\,\raisebox{.15ex}{/}
\hspace*{-\bredde} #1\else$\,\raisebox{.15ex}{/}\hspace*{-\bredde} #1$\fi}
\newcommand{\be}{\begin{equation}}
\newcommand{\ee}{\end{equation}}
\newcommand{\bea}{\begin{eqnarray}}
\newcommand{\eea}{\end{eqnarray}}
\newcommand{\nn}{\nonumber}
\newcommand{\al}{\alpha}

\newcommand{\ka}{\kappa}

\newcommand{\sig}{\sigma}

\newcommand{\sect}[1]{\setcounter{equation}{0}\section{#1}}

\def\tr{{\mbox{tr}}}
\def\re{{\Re\mbox{e}}}
\def\im{{\Im\mbox{m}}}
\def\Jd{J^\dagger}

\begin{document}
\topmargin -1.4cm
\oddsidemargin -0.8cm
\evensidemargin -0.8cm
\title{\Large{{\bf 
Characteristic Polynomials of Complex Random Matrix Models }}}

\vspace{1.5cm}
\author{~\\{\sc G. Akemann} and {\sc G. Vernizzi}\\~\\
Service de Physique Th\'eorique, CEA/DSM/SPhT Saclay\\
Unit\'e de recherche associ\'ee au CNRS\\
F-91191 Gif-sur-Yvette C\'edex, France
}

\date{}
\maketitle
\vfill
\begin{abstract}
We calculate the expectation value of an arbitrary product of
characteristic polynomials of complex random matrices and their
hermitian conjugates.  Using the technique of orthogonal polynomials
in the complex plane our result can be written in terms of a
determinant containing these polynomials and their kernel.  It
generalizes the known expression for hermitian matrices and it also
provides a generalization of the Christoffel formula to the complex
plane.  The derivation we present holds for complex matrix models with
a general weight function at finite-$N$, where $N$ is the size of the
matrix.  We give some explicit examples at finite-$N$ for specific
weight functions.  The characteristic polynomials in the large-$N$
limit at weak and strong non-hermiticity follow easily and they are
universal in the weak limit.  We also comment on the issue of the BMN
large-$N$ limit.
\end{abstract}
\vfill

\begin{flushleft}
SPhT T02/175\\
hep-th/0212051
\end{flushleft}
\thispagestyle{empty}
\newpage

\renewcommand{\thefootnote}{\arabic{footnote}}
\setcounter{footnote}{0}

\sect{Introduction}\label{intro}

The study of the spectral statistical properties of random matrix
models naturally leads to consider characteristic polynomials of
random matrices.  Because of the wide applicability of random
matrices, the analysis of characteristic polynomials has been
approached from different fields of physics and mathematics, such as
quantum chaos \cite{AS}, Quantum Chromodynamics (QCD) \cite{Jac} or
the study of the statistical distribution of the zeros of the Riemann
zeta-function \cite{KS} to name a few of them.

The expectation value of ratios of random matrix determinants
including sources has been introduced in the supersymmetric approach
to random matrix models, as it can be used as a generating functional
for resolvents and thus for all eigenvalue correlation functions. In
the supersymmetric approach the determinants are expressed as
integrals over fermionic and bosonic variables which can then be
evaluated at finite-$N$ or large-$N$, where $N$ is the size of the
matrix. For a review of this approach we refer to \cite{GMW}.
Recently the Ingham-Siegel integral has been used as an alternative to
the Hubbard-Stratonovich transformation \cite{yanIS}.  Another
approach to calculate ratios of characteristic polynomials is the
replica method, where new exact results have been obtained very
recently by \cite{Eugene} and \cite{KimJac} (see also \cite{FSA}).  A
third approach which we will adopt in the following, is the method of
orthogonal polynomials. Arbitrary ratios of characteristic polynomials
have been calculated by means of this method in \cite{FSA,FS} and a
rigorous proof of their universality has been provided in
\cite{FSuniv}.

In this article we will focus on products of characteristic
polynomials. In principle they contain all informations about the
eigenvalues.  When the matrices are hermitian, products of
characteristic polynomials exhibit an interesting duality property,
which interchanges the size of the matrix with the number of
determinants \cite{MN,FW}.  Moreover, in the application to QCD they
enjoy a direct interpretation as partition functions including quark
masses \cite{Jac}.  They also give the orthogonal polynomials and the
kernel with respect to a weight function which includes such quark
mass terms.  Consequently all massive correlation functions can be
obtained from them \cite{AD}\footnote{Here we are not considering the
relation to finite volume partition functions of QCD, but only to
random matrix models partition functions.}.  In fact many of the
results for characteristic polynomials have been first obtained in the
QCD-related literature in the large-$N$ limit, without explicitly
calling them ``products of characteristic polynomials'' but ``massive
partition functions''. For the unitary ensemble of random matrices
they were first calculated in \cite{DN2} including a universality
proof.  For the chiral unitary ensemble the partition functions were
derived in \cite{GWW} for finite-$N$ and infinite-$N$ and shown to be
universal \cite{Poul}. For the other symmetry classes of chiral
orthogonal and symplectic ensembles the massive partition functions
were derived in \cite{NN1} and \cite{AK}, and their universality has
been proved \cite{AK}.  The corresponding nonchiral results were
obtained in \cite{NN2}.  Results for finite-$N$ and infinite-$N$ were
independently found for all three nonchiral symmetry classes in
\cite{BH} and \cite{MN}.

Up to now such results were little explored for random matrix models
with complex matrices having complex eigenvalues. Our aim is to
provide an answer at finite-$N$ and large-$N$ for products of
characteristic polynomials where one has to distinguish between the
matrix and its hermitian conjugate now.  Several developments in
various fields have renewed the interest in complex matrix
models. Such models have been related for example to the
Saffman-Taylor instability on the interface of two-dimensional fluids
\cite{ABWZ}, and correlation functions of complex eigenvalues at large
distance were calculated in
\cite{WZ}. The correspondence
between supersymmetric Yang-Mills theory and string theory has also
renewed the interest in expectation values of traces of complex
matrices \cite{plefki}. There, the complex matrix model belongs to the
Ginibre ensemble and is a tool for computing operators in a peculiar
large-$N$ limit, the so-called Berenstein-Maldacena-Nastase (BMN)
limit \cite{BMN}.  The computational techniques range from
combinatorics, character expansion, orthogonal polynomials (as in the
original approach of Ginibre \cite{Gin}), up to large-$N$ loop
equation techniques \cite{EK} or collective field theory \cite{KJR}.

Matrices with complex eigenvalues occur also in QCD when a chemical
potential for the quarks is introduced \cite{Steph}.  As QCD belongs
to the chiral symmetry class, a chiral complex eigenvalue model has
been introduced and solved in \cite{A02}.  The nonchiral complex model
with unitary symmetry, corresponding to three-dimensional QCD, was
previously studied in \cite{A01}. There, the massive partition
functions as well as all correlation functions were determined under
the hypothesis that the mass terms provide a non-negative definite
measure.  Here, we will generalize these results to arbitrary products
of characteristic polynomials, without any restriction on the sources.

The technique we use in the present article is based on general
orthogonal polynomials in the complex plane. Many important properties
that usually hold for such polynomials on the real axis (e.g. the
existence of a three-step recursion relation or the
Christoffel-Darboux formula) are no longer valid in the complex plane
in general.  Our results will rely solely on the existence of the
orthogonal polynomials.

The article is organized as follows. After some definitions we first
state and prove our main result in Section \ref{Prodchp} for any
product of characteristic polynomials of complex matrices and their
hermitian conjugate. As a nice consequence we also find a simple
determinant expression for products of characteristic polynomials of
hermitian matrices, when the number of factors is odd.  In Section
\ref{Appl} we give an interpretation of our findings. In some special
case they give massive orthogonal polynomials and their kernel, which
leads to all massive eigenvalue correlation functions. We also make
contact to previous results in \cite{A01}.  In Section \ref{finiteN}
we provide some explicit examples of orthogonal polynomials in the
complex plane at finite-$N$.  We will also extend the duality
\cite{MN,FW} between products of characteristic polynomials of
different matrix size to complex matrices.  Section \ref{largeN} is
devoted to the microscopic large-$N$ limit. The known results for the
asymptotics of orthogonal polynomials and their kernel at weak and
strong non-hermiticity directly apply to the asymptotics for the
characteristic polynomials. Also, the proof of universality at weak
non-hermiticity \cite{A02II} can be extended to our case.  We note
however that at strong non-hermiticity only the kernel has a smooth
microscopic large-$N$ limit (and not the polynomials).  Before the
conclusions we also briefly comment in Section \ref{BMNlim} on the
informations about the BMN large-$N$ limit that can be extracted from
the products of characteristic polynomials.

\sect{Products of characteristic polynomials}\label{Prodchp}

We define our complex matrix model partition function and the
corresponding expectation values by
\bea
\tilde{\cal Z}_N &\equiv& \int[dJ\,d\Jd] \ w(J,\Jd) \, , \quad 
\label{Z} \\
\langle\, {\cal O}\, \rangle_N &\equiv& \frac{1}{\tilde{\cal Z}_N}
 \int[dJ\,d\Jd] \ w(J,\Jd)\ {\cal O}(J,\Jd)\ \, ,
\eea
respectively. Here $J$ is a general complex matrix of size $N\times
N$ and we integrate over all independent matrix elements.  We restrict
ourselves to real non-negative weight functions $w(J,\Jd)$.  In
general a complex matrix can be diagonalized by using {\it two}
unitary matrices.  Alternatively it is possible to transform it into a
triangular form by a single unitary transformation, the Schur
decomposition
\be
J\ =\ U(Z\ +\ R )U^{-1}\ ,\ \  
Z=\mbox{diag}(z_1,\ldots,z_N)\, , \ \ U \in  \mbox{U}(N), \ z_i\in\mathbb{C} \ ,
\label{Schur}
\ee
where $R$ is a strictly upper triangular matrix with complex entries.
We assume that the above partition function can be written solely in
terms of the complex eigenvalues $z_i$, $i=1,\ldots,N$ and that the
weight $w(J,\Jd)$ factorizes. We thus can write
\be
\tilde{\cal Z}_N \ =\ C(N)\int_D\prod_{i=1}^N \left( d^2z_i\ 
          w(z_i,\bar{z}_i)\right)
\left|\Delta_N(z)\right|^2\ 
\equiv \  C(N)\ {\cal Z}_N \ \ ,
\label{Zev}
\ee
where we have used that the Jacobian of the transformation
(\ref{Schur}) yields the Vandermonde determinant denoted by
\be
\Delta_N(z)\ \equiv\ \prod_{i>j}^N(z_i-z_j)\ =\ 
\det_{1\leq i,j\leq N}[z_i^{j-1}] \ .
\label{Vander}
\ee
The constant $C(N)$ contains the integral over the unitary group
$U(N)$ as well as the integral over the upper triangular matrix $R$.
The latter drops out also in other expectation values, as for example
in the determinants we consider below.  In eq. (\ref{Zev}) $D$ denotes
the domain in the complex plane over which we integrate the complex
eigenvalues.  Examples for weight functions satisfying the above
conditions are\footnote{In our conventions we will not put a factor of
$N$ into the exponent in order to have exact results for
finite-$N$. When taking the large-$N$ limit it can be reestablished by
scaling $z_i\to \sqrt{N}z_i$ and by suitably rescaling the coupling
constants in the potential $V$.}
\be
w(J,\Jd) \ = \ \exp[\tr V(J,\Jd)], \ \ \mbox{with} \ 
V(J,\Jd) \ =\ g J\Jd + V(J) + [V(J)]^{\dagger} \ .
\label{expweight}
\ee
Here $V(J)$ can be taken for example as an arbitrary complex
polynomial and $g$ is a real coupling constant. The domain $D$ can be
the full complex plane or any bounded domain, depending on the weight
function. For instance if
\be
w(J,\Jd) \ =\ \delta\left(\tr (J\Jd)-r^2\right) \, , \quad \mbox{or} \quad
w(J,\Jd) \ =\ \theta\left(\tr (J\Jd)-r^2\right) \, ,
\label{deltaweight}
\ee
then the domain is obviously a circle or a disk 
of radius $r$ around the origin in the complex plane, 
respectively.

If the weight function $w(z,\bar{z})$ is such that all moments exist,
$\int_D d^2z \ w(z,\bar{z}) \ |z|^{2n}< \infty$, $\forall n \in
\mathbb{N}$, then the orthogonal polynomials $\tilde{P}_n(z)$
\be
\int_D d^2z \ w(z,\bar{z})\tilde{P}_k(z) \ 
\overline{\tilde{P}_l(z)} 
= h_k\delta_{kl}  \ , 
\label{OPdef}
\ee
exist as well and are uniquely determined by the monic normalization
$\tilde{P}_k(z)=z^k+O(z^{k-1})$ with norm $h_k$.  Note that the
coefficients of $\tilde{P}_l(z)$ are complex in general.  However if
both $w(z,\bar{z})$ and the domain $D$ are symmetric under the
exchange $z \leftrightarrow \bar{z}$ then $\tilde{P}_l(z)$ can be
chosen with real coefficients.  The theory of orthogonal polynomials
in the complex plane (and their asymptotics) is a well-established
subject of mathematics; for further details, we refer the interested
reader to the mathematical literature (e.g. \cite{stahl}).

We can easily express the eigenvalue partition function
eq. (\ref{Zev}) in terms of the norms $h_k$ by replacing the
Vandermonde determinant eq. (\ref{Vander}) with a determinant of the
monic polynomials $\tilde{P}_k(z)$ and using the orthogonality
relation (\ref{OPdef}):
\be
{\cal Z}_N \ =\ N! \prod_{k=0}^{N-1} h_k \ .
\label{Zdef}
\ee
In order to state our results we also introduce the ortho{\it normal}
polynomials
\be
P_k(z)\ \equiv\ h_k^{-\frac12} \tilde{P}_k(z) \, .
\label{ONdef}
\ee
With these preliminaries we can turn to the main result of this article.
 
\ \\

\noindent
\underline{\sc Theorem}: 
Let $\{v_i;\, i=1,\ldots,K\}$ and $\{u_i;\, i=1,\ldots,L\}$ be two
sets of complex numbers which are pairwise distinct among each set.
Without loss of generality we assume $K\geq L$, where the empty set
with $L=0$ is permitted as well.  Together with the definitions and
restrictions on the weight function made above, the following
statement holds\footnote{The following notation is understood:
$\Delta_0(x)=\Delta_1(x)=1$ and $\prod_{i=N}^{N-1}h_i=1$ }:
\be
\left\langle  \prod_{i=1}^K\det[v_i-J]\ \prod_{j=1}^L\det[\bar{u}_j-\Jd]
\right\rangle_N
\ =\  \frac{\prod_{i=N}^{N+K-1}h_i^{\frac12}\ 
\prod_{j=N}^{N+L-1} h_j^{\frac12}}{\Delta_K(v)\ \Delta_L(\bar{u})}
\det_{1\leq l,m\leq K}[\ {\cal B}(v_l,\bar{u}_m)\ ] \ ,
\label{Th}
\ee
where we have defined:
\be
{\cal B}(v_l,\bar{u}_m) \ \equiv\ \left\{ 
\begin{array}{ccl}
\sum_{i=0}^{N+L-1}P_i(v_l)\overline{P_i(u_m)} & \mbox{for} & m=1,\ldots,L \\
&&\\
P_{N+m-1}(v_l)                           & \mbox{for} & m=L+1,\ldots,K \\
\end{array}
\right. \ .
\label{Ddef}
\ee

\ \\

Before presenting the proof of the Theorem let us make a few remarks.
The reason why we restrict ourselves to the case $K\geq L$ is that the
other case $K\leq L$ can be obtained simply by complex conjugating the
above equations. The restriction to pairwise distinct sets of
parameters $v_i$ or $u_i$ can also be lifted. In the limit $v_i \to
v_j$ the Vandermonde determinant in the denominator vanishes as well
as the determinant in the numerator.  Hence in this limit the row
containing $v_j$ has to be differentiated with respect to $v_j$ and
then we set $v_j=v_i$ there. The same procedure can be iterated when
$v_i$ is degenerate of order $m$. The result is given by substituting
the $k$-th row of eq. (\ref{Ddef}) for $k=1,\ldots,m$ with its
$(k-1)$-th derivative at $v_i$ (and dropping the corresponding
arguments in the Vandermonde determinants in the denominator). The
same argument applies for the case with degenerate $\bar{u}_i$.  We
will give an example for $L=0$ in eq. (\ref{deg}).

We also wish to mention that the object occurring in the first $L$
columns of the matrix ${\cal B}(v_l,\bar{u}_m)$ is, apart from the
weight factors, nothing else than the kernel of orthogonal polynomials
defined as:
\be
K_{N}(v,\bar{u})\ \equiv\ [w(v,\bar{v})w(u,\bar{u})]^{\frac12}
\sum_{i=0}^{N-1}P_i(v)\overline{P_i(u)}
\ \equiv\  [w(v,\bar{v})w(u,\bar{u})]^{\frac12} \ \ka_{N}(v,\bar{u})
\ .
\label{Kerdef}
\ee
Here we have also explicitly defined the bare kernel
$\ka_{N}(v,\bar{u})$ as it occurs in the Theorem and subsequent
formulas.  We note that this kernel cannot be written in terms of
polynomials of order $N$ and $N-1$ alone because for orthogonal
polynomials in the complex plane the analog of the Christoffel-Darboux
formula, eq. (\ref{CD}) below, does not hold in general (for an
example see \cite{A02II}).

\ \\

\noindent
\underline{\sc Proof}: 
The proof will be done in two steps. First we will prove the case
$K=L$ for arbitrary $L$. Then we consider the case $K>L$ using
induction.  For $K=L$ we have:
\bea
\lefteqn{\left\langle  \prod_{i=1}^L\det[v_i-J]\ 
\prod_{j=1}^L\det[\bar{u}_j-\Jd]
\right\rangle_N = {} } \nn\\
&&=\ \frac{1}{\cal Z_N} \int_D \prod_{i=1}^N\left( d^2z_i\ 
w(z_i,\bar{z}_i)\ \prod_{j=1}^L((v_j-z_i)(\bar{u}_j-\bar{z_i}))\
\right) \Delta_N(z)\Delta_N(\bar{z})\nn\\
&&=\ 
 \frac{1}{{\cal Z}_N} \int_D \left( \prod_{i=1}^N d^2z_i\ w(z_i,\bar{z}_i)\ 
\right)\frac{\Delta_{N+L}(z_1,\ldots,z_n,v_1,\ldots,v_L)}{
\Delta_{L}(v_1,\ldots,v_L)}
\frac{\Delta_{N+L}(\bar{z}_1,\ldots,\bar{z}_n,\bar{u}_1,\ldots,\bar{u}_L)}{
\Delta_{L}(\bar{u}_1,\ldots,\bar{u}_L)}\nn\\
&&=\  \frac{1}{{\cal Z}_N \Delta_{L}(v) \Delta_{L}(\bar{u})}
 \int_D \prod_{i=1}^N\left( d^2z_i\ w(z_i,\bar{z}_i)\ \right)
\det_{1\leq i,j\leq N+L}[\zeta_i^{j-1}]
\det_{1\leq i,j\leq N+L}[\bar{\eta}_i^{\ j-1}] \, , \  \mbox{with} \  
\left\{
\begin{array}{l}
\zeta\equiv \{z,v \} \\
\eta\equiv \{ z,u \} 
\end{array}
\right.
\nn\\
&&=\  \frac{\prod_{i=0}^{N+L-1} h_i}{{\cal Z}_N\Delta_{L}(v)
\Delta_{L}(\bar{u})} \int_D \prod_{i=1}^N
\left( d^2z_i\ w(z_i,\bar{z}_i)\  \right)
\det_{1\leq i,j\leq N+L}[P_{j-1}(\zeta_i)]
\det_{1\leq i,j\leq N+L}[\overline{P_{j-1}(\eta_i)}]  \nn\\
&&=\  \frac{\prod_{i=N}^{N+L-1} h_i}{N! \Delta_{L}(v) 
\Delta_{L}(\bar{u})} \int_D \prod_{i=1}^N 
\left( d^2z_i\ w(z_i,\bar{z}_i) \ \right)
\det_{1\leq i,j\leq N+L}[\sum_{r=1}^{N+L} P_{r-1}(\zeta_i) 
\overline{P_{r-1}(\eta_j)}]  \ .
\label{proof1}
\eea
In the second step we have put the products of differences between
$v_j$ and $z_i$ into a larger Vandermonde determinant of size $N+L$
and between $\bar{u}_j$ and $\bar{z}_i$ into a complex conjugated one.
The additional factors of differences between $v_j$'s and
$\bar{u}_j$'s have been divided out in terms of smaller Vandermonde
determinants of size $L$.  The Vandermonde determinant can be written
as a determinant of powers, eq.  (\ref{Vander}), where we have
introduced a larger set of variables $\zeta_{1,\ldots,N+L}$ and
$\eta_{1,\ldots,N+L}$.  The properties of determinants allow to
replace the powers $\zeta^{j-1}_i$ first by the monic polynomials
$\tilde{P}_{j-1}(\zeta_i)$ and then by the orthonormal ones times
their norms, which have been taken out of the determinants. They
partially cancel the normalization ${\cal Z}_N$ from eq.
(\ref{Zdef}). We then apply that $\det[A]\det[B]=\det[AB^T]$ and thus
obtain a single determinant of the bare kernel eq. (\ref{Kerdef}).
Now we make use of Theorem 5.2.1 in \cite{Mehta}, which can be applied
as follows. The kernel $K_N(\zeta,\bar{\eta})$ satisfies the
hermiticity condition
$\overline{K_N(\zeta,\bar{\eta})}=K_N(\eta,\bar{\zeta})$ and the chain
relation $\int d^2z\ K_N(\zeta,\bar{z})\ K_N(z,\bar{\eta})
=K_N(\zeta,\bar{\eta})$.  Then
\be 
\int d^2z_n \ \det_{1 \leq i,j \leq n} [ K_N(z_i,\bar{z}_j)]  \ =\ (c-n+1)
\det_{1 \leq i,j \leq n-1}[ K_{N}(z_i,\bar{z}_j)] \, ,
\label{MTh}
\ee
where $c= \int d^2 z K_N(z,\bar{z})=N$.  We can thus successively
integrate out all $z_{1,\ldots,N}$ using eq. (\ref{MTh}).  The
successive use produces a factor $N!$ and we thus arrive precisely at
eq. (\ref{Th}) for $K=L$, with the matrix ${\cal B}$ only consisting
of the bare kernels.

Next we prove eq. (\ref{Th}) for $K>L$ by induction in $K-L$.  The
starting point of induction is already satisfied for $K=L$.  For the
induction step $K\to K+1$ we proceed as follows:
\bea
\lefteqn{\left\langle  \prod_{i=1}^{K+1}\det[v_i-J]\ 
\prod_{j=1}^L\det[\bar{u}_j-\Jd]
\right\rangle_N = {} } \nn\\
&&=\  \frac{1}{{\cal Z}_N\Delta_{K+1}(v)
\Delta_{L}(\bar{u})} \int_D \prod_{i=1}^N d^2z_i\ w(z_i,\bar{z}_i) \nn\\
&&\ \times
\sum_{\sig\in\{0,\ldots,N+K\}}(-1)^\sig 
\tilde{P}_{\sig(1)}(z_1)\cdots\tilde{P}_{\sig(N)}(z_N)
\tilde{P}_{\sig(N+1)}(v_1)\cdots\tilde{P}_{\sig(N+K)}(v_K)
\tilde{P}_{\sig(N+K+1)}(v_{K+1})\nn\\
&&\ \times
\sum_{\sig'\in\{0,\ldots,N+L-1\}}(-1)^{\sig'}
\overline{\tilde{P}_{\sig'(1)}(z_1)}\cdots
\overline{\tilde{P}_{\sig'(N)}(z_N)}
\,\overline{\tilde{P}_{\sig'(N+1)}(u_1)}\cdots
\overline{\tilde{P}_{\sig'(N+L)}(u_L)}\nn\\
&&=\ 
\frac{\prod_{i=N}^{N+K}h_i^{\frac12}\ \prod_{j=N}^{N+L-1}h_j^{\frac12}
}{\Delta_{K+1}(v)\ \Delta_L(\bar{u})}
\left( {P}_{N+K}(v_{K+1})
\det_{1\leq l,m\leq K}\left[{\cal B}(v_l,\bar{u}_m)\right] \right. \nn\\
&&\ \  \left. - \sum_{j=1}^K  {P}_{N+K}(v_{j})
\det_{\stackrel{m=1,\ldots,K}{l=1,\ldots,j-1,K+1,j+1,\ldots,K}}\left[
{\cal B}(v_l,\bar{u}_m)
\right]\right) \ .
\label{induction} 
\eea
We repeated the first steps in eq. (\ref{proof1}) until we have the
product of two determinants of monic polynomials and their conjugate,
containing the variables $\{z_{1,\ldots,N},v_{1,\ldots,K+1}\}$ and
$\{z_{1,\ldots,N},\bar{u}_{1,\ldots,L}\}$, respectively.  We have
explicitly spelled out these determinants as sums over permutations.
The index $\sig(N+K+1)$ of the polynomial
$\tilde{P}_{\sig(N+K+1)}(v_{K+1})$ can take values in
$\{0,\ldots,N+K\}$.  For $\sig(N+K+1)=N+K$ the polynomial
$\tilde{P}_{N+K}(v_{K+1})$ multiplies exactly the expectation value
$\langle \prod_{i=1}^K\det[v_i-J] \prod_{j=1}^L
\det[\bar{u}_j-\Jd]\rangle_N$ for which the Theorem holds by using the
assumption of induction. The other values for the index
$\sig(N+K+1)<N+K$ can be obtained by simply exchanging the argument
$v_{K+1}$ with the one of the corresponding polynomial of argument
$v_{j=1,\ldots,K}$. Since this is a pair permutation in $\sig$ it will
produce a sign for all terms.  As a final step, in the last line of
eq. (\ref{induction}) we move the row containing $v_{K+1}$ from the
position $j$ to the lowest row of position $K$, without changing the
order of the other rows in the determinants.  This produces a factor
$(-1)^{K-j}$.  Then we see that the sum of all determinants is just
the Laplace expansion of a matrix of size $K+1$ with respect to the
last column, and that this matrix is nothing else than ${\cal
B}(v_l,\bar{u}_m)$ in eq. (\ref{Ddef}).  We have thus proved the
Theorem for $K+1$.

\ \\

Let us emphasize the Theorem (\ref{Th}) also holds for partition
functions defined as eigenvalue integrals such as eq. (\ref{Zev}) and
which do not necessarily have a representation in terms of complex
matrices.  Examples for such models are the chiral complex matrix
model as recently proposed in \cite{A02} which we will discuss again
in Section \ref{finiteN}.  For this model no explicit matrix
realization is known so far. Another example are weight functions
containing mixed higher powers of $z$ and $\bar{z}$ such as
$w(z,\bar{z})=\exp[-(z\bar{z})^{k\geq2}]$. Such weight functions can
be only written down in terms of complex matrices which are normal,
$[J,\Jd]=0$, as normal matrices can be diagonalized without an upper
triangular matrix as in eq. (\ref{Schur}). However, such terms with
higher powers in $|z|^2$ may be necessary in the weight function to
write down convergent integrals.

After having completed the proof let us compare to the known results
for products of characteristic polynomials of hermitian matrices
$J=H=H^\dagger$. Since in our proof we have only assumed the existence
of an eigenvalue representation for the partition function and the
existence of orthogonal polynomials on some domain $D$, it holds also
for arbitrary hermitian matrix models. We only have to restrict $D$ to
the real line and replace complex integrals $\int d^2z$ by real
integrals $\int dx$. Because of $J=J^\dagger$ the splitting into two
sets of determinants becomes immaterial here.  In particular for $L=0$
we find back the known result (see e.g.  in
\cite{BH,MN,FW})
\be
\left\langle  \prod_{i=1}^K\det[v_i-H]
\right\rangle_N
\ =\  \frac{1}{\Delta_K(v)}
\det_{1\leq l,m\leq K}[\ \tilde{P}_{N+m-1}(v_l) \ ] \ .
\label{ThP}
\ee
The product of $K$ characteristic polynomials is thus given by a
$K\times K$ determinant of the orthogonal polynomials. In
\cite{FSuniv} it was shown that for even $K=2k$ this can be further
simplified to a $k\times k$ determinant.  We obtain the same result
from setting $K=L=k$ in our Theorem $(\ref{Th})$ and using that for
orthogonal polynomials on the real line the Christoffel-Darboux
identity holds,
\be
\sum_{i=0}^{N-1}P_i(x)P_i(y)\ =\ 
\sqrt{\frac{h_{N}}{h_{N-1}}}\frac{P_N(x)P_{N-1}(y)-P_N(y)P_{N-1}(x)}{x-y}\, , 
\label{CD}
\ee
with $x,y$ real numbers\footnote{Formula (\ref{CD}) holds
algebraically also for $x,y \in \mathbb{C}$ and $P_k(x)$ being the
analytic continuation to the complex plane of the polynomials which
are orthonormal on the real line. However we emphasize that in this
case the sum $\sum_{i=0}^{N-1}P_i(x)P_i(y)$ is not a kernel in the
complex plane, because the $P_k$ are not longer orthonormal in the
complex plane (for an explicit example see eq. (\ref{Herm}) ).}.  We
thus have
\bea
\lefteqn{\left\langle  \prod_{i=1}^k\left(\det[v_i-H]\det[u_i-H]\right)
\right\rangle_N = {} }\nn\\
&&=\  \frac{\prod_{i=N}^{N+k-1}h_i}{(h_{N+k-1})^k
\Delta_k(v)\Delta_k(u)}
\det_{1\leq l,m\leq k}\left[ 
\frac{\tilde{P}_{N+k}(v_l)\tilde{P}_{N+k-1}(u_m)-
\tilde{P}_{N+k}(u_m)\tilde{P}_{N+k-1}(v_l)}{v_l-u_m}
\right] \, .
\label{ThK}
\eea
As a simple corollary we can thus show from the Theorem (\ref{Th}) that
also for and odd number $K=2k+1$ of products of characteristic
polynomials the $K\times K$ determinant in eq. (\ref{ThP}) can be
reduced down to size $(k+1)\times(k+1)$.

\ \\

\noindent
\underline{\sc Corollary}: 
Let the definitions (\ref{Z}) - (\ref{Vander}) and (\ref{OPdef}) -
(\ref{ONdef}) be valid as well for hermitian matrices $H$ with real
eigenvalues integrated over a real domain $D$. Take the parameters
(real or complex) $\{v_i;\, i=1,\ldots,k+1\}$ and $\{u_i;\,
i=1,\ldots,k\}$ to be all pairwise distinct. From Theorem (\ref{Th})
the following relation follows:
\be
\left\langle  \prod_{i=1}^{k+1}\det[v_i-H]\ \prod_{j=1}^k\det[u_j-H]
\right\rangle_N
\ =\  \frac{\prod_{i=N}^{N+k-1}h_i}{(h_{N+k-1})^k
\Delta_{k+1}(v)\Delta_{k}(u)}
\det_{1\leq l,m\leq k+1}[\ \mbox{\bf B}(v_l,u_m)\ ] \ ,
\label{ThKP}
\ee
where we have defined
\be
\mbox{\bf B}(v_l,u_m) \ \equiv\ \left\{ 
\begin{array}{ccl}
\frac{\tilde{P}_{N+k}(v_l)\tilde{P}_{N+k-1}(u_m)-
\tilde{P}_{N+k}(u_m)\tilde{P}_{N+k-1}(v_l)}{v_l-u_m}
& \mbox{for} & m=1,\ldots,k \\
&&\\
\tilde{P}_{N+m-1}(v_l)                           & \mbox{for} & m=k+1 \\
\end{array}
\right.\ .
\label{bDdef}
\ee

\ \\
 
It is clear that from Theorem (\ref{Th}) for real hermitian matrices
we can get many equivalent formulas for determinants of size in
between eq. (\ref{ThP}) and eqs. (\ref{ThK}) or (\ref{ThKP})
respectively, by splitting the total number of parameters in different
pairs $K$ and $L$.

\sect{Applications}\label{Appl}

In this Section we give an interpretation for some particular products
of characteristic polynomials we calculated in eq. (\ref{Th}). From
the beginning of our investigations we have assumed that the weight
function has to be non-negative definite on the domain $D$ in order to
make the construction of the polynomials via Gram-Schmidt
possible. The examples given for $w(z,\bar{z})$ in
eqs. (\ref{expweight}) and (\ref{deltaweight}) obviously satisfy this
requirement. We will now use the results from our Theorem to
explicitly construct orthogonal polynomials for more complicated
weight functions containing determinants or in other word weights with
products of eigenvalues as prefactors.

From the theory of orthogonal polynomials defined on the real line the
following is known. Choosing any parameter $v_j\in \mathbb{R}$ in
eq. (\ref{ThP}) for hermitian matrices the right hand side can be
interpreted as a polynomial in $v_j$ of order $N$, $\tilde{P}_N(v_j)$,
which is orthogonal with respect to the weight $w^{(K-1)}(x)\equiv
w(x)\prod_{i\neq j}(v_i-x)$.  This holds provided that the new weight
is non-negative definite on the domain $D$. Note that under this
condition, eq. (\ref{ThP}) immediately follows from the Christoffel
Theorem \cite{szego}.  An example is when $D \subseteq \mathbb{R}$ and
all $v_i$ are purely imaginary, coming in complex conjugate
pairs. This would correspond to the weight for the random matrix
partition function of QCD in three dimensions.

In order to make such an interpretation of the characteristic
polynomial in the complex plane possible we have to choose $K=L+1$ in
eq. (\ref{Th}) and to take all $u_i=v_i$, $i=1,\ldots,L$.  It is easy
to convince oneself by going to an eigenvalue basis that the following
matrix representation holds:
\be
\tilde{P}^{(L)}_N(z) \ \sim\ 
\left\langle  \det[z-J]
\prod_{i=1}^L\left(\det[v_i-J]\det[\bar{v}_i-\Jd]\right) \ 
\right\rangle_N  \, .
\label{LOP}
\ee
This gives a 
polynomial orthogonal with respect to the non-negative definite weight 
\be
w^{(L)}(J,\Jd)\ \equiv\ w(J,\Jd)
\prod_{i=1}^L\left(\det[v_i-J]\det[\bar{v}_i-\Jd]\right) \ ,
\label{Lweight} 
\ee
and the complex conjugate polynomial $\overline{\tilde{P}^{(L)}_N(z)}$
is simply obtained by taking the hermitian conjugate of
eq. (\ref{LOP}).  Please note the exact analogy with a similar formula
for orthogonal polynomials on the real line (see e.g. \cite{szego}).
In order to satisfy the normalization
$\tilde{P}^{(L)}_N(z)=z^N+\ldots$ we define \be
\tilde{P}^{(L)}_N(z) \ \equiv\ \frac{\left\langle  \det[z-J]
\prod_{i=1}^L\left(\det[v_i-J]\det[\bar{v}_i-\Jd]\right)
\right\rangle_N}{\left\langle 
\prod_{i=1}^L\left(\det[v_i-J]\det[\bar{v}_i-\Jd]\right)
\right\rangle_N} \ .
\label{LOPdef}
\ee
Let us give an example for $L=1$: 
\be
\tilde{P}^{(1)}_N(z) \ =\  \frac{\ka_{N+1}(z,\bar{v})
\tilde{P}_{N+1}(v)-\ka_{N+1}(v,\bar{v})\tilde{P}_{N+1}(z)}{(v-z)\ 
\ka_{N+1}(v,\bar{v})} \ ,
\label{OPunL=1}
\ee
where we have used the bare kernel $\ka_{N+1}(v,\bar{u})$
eq. (\ref{Kerdef}) and the superscript $(L=0)$ has been dropped.  We
can also calculate the norm of the polynomial given by
\be
h^{(1)}_{N}\ =\ h_{N+1}\ \frac{\ka_{N+2}(v,\bar{v})}{\ka_{N+1}(v,\bar{v})}\ ,
\ee
and we obtain for the orthonormal polynomials 
\be
P^{(1)}_N(z) \ =\ \frac{\ka_{N+1}(z,\bar{v})
P_{N+1}(v)-\ka_{N+1}(v,\bar{v})P_{N+1}(z)}{(v-z)
\sqrt{\ka_{N+1}(v,\bar{v})\ka_{N+2}(v,\bar{v})}}  \ .
\label{OPL=1}
\ee
We have not been able to prove a similar expression for the
orthonormal polynomials at general $L$. We expect to obtain the
following result for the norm of $\tilde{P}^{(L)}_N(z)$:
\be
h^{(L)}_{N}\ =\ h_{N+L}\ 
\frac{\det_{i,j=1,\ldots,L}[\ka_{N+L+1}(v_i,\bar{v}_j)]}{
\det_{i,j=1,\ldots,L}[\ka_{N+L}(v_i,\bar{v}_j)}\ ,
\label{Lnorm}
\ee
which we have checked for $L=1,2,3,4,5$. This would give us $P^{(L)}_N(z)$
in a form very similar to \cite{FG}.

After having determined the orthogonal polynomials for weight
functions $w^{(L)}(J,\Jd)$ we can use them to obtain the respective
kernel as defined in eq (\ref{Kerdef}):
\be
K^{(L)}_{N}(z,\bar{u})\ =\ [w^{(L)}(z,\bar{z})w^{(L)}(u,\bar{u})]^{\frac12}
\sum_{i=0}^{N-1}P^{(L)}_i(z)\overline{P^{(L)}_i(u)}
\ .
\label{Lkerdef}
\ee
While we have an explicit expression for $L=1$ in eq. (\ref{OPL=1}) we
would still have to calculate the norms $h_N^{(L)}$ of the polynomials
$\tilde{P}^{(L)}_N(z)$ which are of the conjectured form
eq. (\ref{Lnorm}).  However, we can also directly read off the kernel
from the Theorem (\ref{Th}).  It is known that not only the
polynomials have a matrix representation as in eq. (\ref{LOPdef}) but
also the kernel itself. Choosing the same weight eq. (\ref{Lweight})
one has to insert two more determinants instead of only one as for the
polynomials:
\be
K^{(L)}_{N}(z,\bar{u})\ =\ [w^{(L)}(z,\bar{z})w^{(L)}(u,\bar{u})]^{\frac12}
\frac{\left\langle  \det[z-J]\det[\bar{u}-\Jd]\ 
\prod_{i=1}^L\left(\det[v_i-J]\det[\bar{v}_i-\Jd]\right)
\right\rangle_{N-1}}{h_{N}\ \left\langle 
\prod_{i=1}^L\left(\det[v_i-J]\det[\bar{v}_i-\Jd]\right)
\right\rangle_{N}}.
\label{Lker}
\ee
This can be shown along the same lines as in \cite{PZJ} where the same
statement was made for hermitian matrices.  Note the different matrix
size of the expectation value for numerator and denominator. We also
give the example with $L=1$ explicitly as it follows from
eq. (\ref{Th}):
\be
K^{(1)}_{N}(z,\bar{u})\ =\ [w^{(1)}(z,\bar{z})w^{(1)}(u,\bar{u})]^{\frac12}\ 
\frac{\left|
\begin{array}{ccc}
\ka_{N+1}(z,\bar{u}) & & \ka_{N+1}(z,\bar{v})\\
\ka_{N+1}(v,\bar{u}) & & \ka_{N+1}(v,\bar{v})
\end{array}
\right|}{(v-z)(\bar{v}-\bar{u})\ \ka_{N+1}(v,\bar{v})}\ .
\label{kerL=1}
\ee
While eqs. (\ref{Lkerdef}) and (\ref{Lker}) are equivalent by
definition, for general $L$ it is easy to see that already for $L=1$
this yields a highly nontrivial identity, comparing eq. (\ref{kerL=1})
and eq. (\ref{Lkerdef}) with the polynomials eq. (\ref{OPL=1})
inserted.  From a practical point of view the second form
eq. (\ref{Lker}) which can be directly deduced from the Theorem
(\ref{Th}) is much more convenient as it is only a ratio of two
determinants.  Compared to this in eq. (\ref{Lkerdef}) we have to
perform a sum over ratios of determinants of growing size.

Usually the aim of calculating orthogonal polynomials and their kernel
is to evaluate correlation functions of eigenvalues defined as
\be
R^{(L)}_N(z_1,\ldots,z_k)\ \equiv\ \frac{1}{{\cal Z}_N}
\frac{N!}{(N-k)!}
\int_D\prod_{l=k+1}^N \!d^2z_l\ 
\prod_{i=1}^N w^{(L)}(z_i,\bar{z}_i)\ \left|\Delta_N(z)\right|^2\ .
\label{Rkdef}
\ee
Using standard orthogonal polynomials techniques \cite{Mehta} they are
given by
\be
R^{(L)}_N(z_1,\ldots,z_k)\ = \ \det_{i,j=1,\ldots,k}
\left[K^{(L)}_{N}(z_i,\bar{z}_j)\right]\ .
\label{MM}
\ee
Therefore from eq. (\ref{Lker}) we can immediately read off all
$k$-point correlation functions. They are given by a $k\times k$
determinant of the kernel $K^{(L)}_{N}(z_i,\bar{z}_j)$ which is itself
again a ratio of determinants of size $(L+1)\times(L+1)$ over $L\times
L$.  There is an even more compact expression for the $k$-point
functions which has been introduced in \cite{A01}. It uses the fact
that the correlation functions with weight $w^{(L)}(J,\Jd)$ from
eq. (\ref{Lweight}) can be entirely expressed in terms of higher
$n$-point correlation functions with weight $w(J,\Jd)$ at $L=0$
\footnote{Note the typo in the index of the denominator in eq. (2.12)
of ref. \cite{A01}.}:
\be
R^{(L)}_N(z_1,\ldots,z_k)\ = \ 
\frac{R^{(0)}_{N+L}(z_1,\ldots,z_k,v_1,\ldots,v_L)}{
R^{(0)}_{N+L}(v_1,\ldots,v_L)}\ .
\label{master}
\ee
This result reduces the $k$-point correlation function to a single
ratio of determinants of size $(k+L)\times(k+L)$ over $L\times L$.
While for the 1-point function $R_N^{(L)}(z)=K_N^{(L)}(z,\bar{z})$ it
is easy to see that eq. (\ref{master}) and eq. (\ref{MM}) together
with eq. (\ref{Lker}) agree perfectly, for higher $(k>2)$-point
functions their equivalence requires highly nontrivial identities
among determinants.  The fact that several different (but equivalent)
determinant formulations exist, depending on the way they are derived,
is a common phenomenon also for correlations functions of hermitian
matrices.  In some cases their equivalence can be directly shown
\cite{ADIII} (see also \cite{Harry}).

\sect{Explicit results for finite-$N$}\label{finiteN}

In the first subsection we will give some explicit examples for weight
functions and their corresponding orthogonal polynomials and kernel at
finite-$N$.  These are the objects that then have to be inserted into
Theorem eq.  (\ref{Th}).  In particular our second example will be
studied also in Section \ref{largeN}, when taking the large-$N$ limit.
In the second subsection we derive a finite-$N$ duality relation
between characteristic polynomials of different matrix size.

\subsection{Examples for orthogonal polynomials in the complex plane}

We start with the Ginibre ensemble where the weight function is
defined on the full complex plane $D=\mathbb{C}$ as
\be
w_G(J,\Jd)\ \equiv\ \frac{1}{\pi} \exp[-\tr J\Jd] \ =\ 
\frac{1}{\pi} \prod_{i=1}^N \exp[-z_i\bar{z_i}] \ .
\label{wGin}
\ee
As one can easily convince oneself the orthogonal polynomials are
monic powers,
\be
\tilde{P}_j(z)\ =\ z^j\ , \ \ h_j=j!\ .
\label{Gin}
\ee
We thus have from the orthonormal polynomials
$P_j(z)=z^j(j!)^{-\frac12}$ a simple expression for the kernel
\be
\ka_{N}(v,\bar{u})\ =\ 
\sum_{i=0}^{N-1}\frac{(v\bar{u})^i}{i\,!}=\exp[v\bar{u}]
\frac{\Gamma(N,v\bar{u})}{\Gamma(N)}\, ,
\label{kerGin}
\ee
where $\Gamma(a,z)= \int_{z}^{\infty}\, dt \, t^{a-1} e^{-t}$ is the
incomplete gamma function. As a consequence of the results in Section
\ref{Appl} we can also immediately read off explicitly the polynomials
for the weight function $w_G^{(L)}(z,\bar{z})=\prod_{i=1}^L
\det[(v_i-z)(\bar{v}_i-\bar{z})]\exp[-z\bar{z}]$ from eq. (\ref{LOPdef}). For 
$L=1$ they are given by:
\bea
\tilde{P}^{(1)}_N(z) &=& \frac{v^{N+1}\sum_{i=0}^N\frac{(z\bar{v})^i}{h_i}
- z^{N+1}
\sum_{i=0}^N\frac{(v\bar{v})^i}{h_i}}{(v-z)\sum_{i=0}^N\frac{(v\bar{v})^i}{
h_i}}\nn\\
&=& v^N \sum_{j=0}^N \left(\frac{z}{v}\right)^j\frac{\ka_{j+1}(v,\bar{v})}{
\ka_{N+1}(v,\bar{v})}\ .
\label{LGin}
\eea
Note that eq. (\ref{LGin}) holds for any $h_i$, and not just for
$h_i=i!$.  Moreover note that we started from polynomials for $L=0$
with real coefficients eq. (\ref{Gin}) while those of the new
polynomials with $L=1$ are now complex.  This example shows a
remarkable fact.  It is well-known that when $D$ is a real interval
there is a correspondence between the distribution of the eigenvalues
and the distribution of the zeros of the orthogonal
polynomials. However in the complex plane such a correspondence breaks
down. For instance, eq. (\ref{Gin}) shows that all the zeros are at
the origin $z=0$, whereas the spectral density spreads into the full
complex plane.

The following example is for the weight 
\be
w_H(J,\Jd)\ \equiv\ \exp\left[-\frac{1}{1-\tau^2}\tr\left( J\Jd
\ -\ \frac{\tau}{2}(J^2+J^{\dagger\,2})\right)\right] \ , 
\ \ \tau\in\ [0,1]\ ,
\label{wHerm}
\ee
which is again defined on $D=\mathbb{C}$.  Here the complex matrix is
parameterized as $J\equiv H+iA\sqrt{\frac{1-\tau}{1+\tau}}$ and $H$
and $A$ are hermitian matrices with equal Gaussian weight. The
parameter $\tau$ measures the degree of non-hermiticity, a concept
which was introduced in \cite{FKS}. In particular this allows to take
the hermitian limit by sending $\tau\to1$.  In this limit
$\lim_{\tau\to1}w_H(J,\Jd)\sim \delta(\im J)\exp[-\tr(\re J)^2]$.
This possibility makes the difference between the complex matrix model
and two-matrix models particularly transparent. In the former we have
introduced polynomials of a complex variable $P_j(z)$ as well as their
complex conjugate. While we can smoothly take the limit of $z$
becoming real here, this is not possible in the two-matrix model
solved by the method of bi-orthogonal polynomials.  The orthogonal
polynomials for the weight in eq. (\ref{Herm}) can be shown to be
Hermite polynomials \cite{PdF},
\be
\tilde{P}_j(z)\ =\ 
\left(\frac{\tau}{2}\right)^{\frac{j}{2}} 
H_j\left(\frac{z}{\sqrt{2\tau}}\right)
\ ,\ \ h_j=\pi j!\,\sqrt{1-\tau^2}\ .
\label{Herm}
\ee
Using that the Hermite polynomials have real coefficients their kernel
is given by
\be
\ka_N(z,\bar{u})\ =
\frac{1}{\pi\,\sqrt{1-\tau^2}}
\sum_{j=0}^{N-1}\frac{1}{j!} \left(\frac{\tau}{2}\right)^j 
H_j\left(\frac{z}{\sqrt{2\tau}}\right)H_j\left(\frac{\bar{u}}{
\sqrt{2\tau}}\right) \, .
\label{kerHerm}
\ee

An other example is for a bounded domain $D$ 
in the complex plane, namely the 
disk $D_r=\{z \in \mathbb{C}: |z|<r\}$ or the circle $C_r=\{z \in 
\mathbb{C}: |z|=r \}$. In both cases the orthogonal polynomials with 
respect to the flat measure $w(z,\bar{z})=1$ are again given 
by monic powers but with different normalization coefficients,
\bea
\tilde{P}_j(z)=z^j\, , \quad && h_j=\frac{\pi}{j+1}\, r^{2 j+2}  
\, , \quad  \mbox{in $D_r$} \, , \label{Disk} \\
\tilde{P}_j(z)=z^j\, , \quad && h_j=2 \pi\, r^{2 j+1}  \, , \quad 
\mbox{in $C_r$}\ ,
\label{Circle}
\eea
and the bare kernels read
\bea
\ka_{N}(v, \bar{u})&&=\frac{1}{\pi r^2}\sum_{j=0}^{N-1} (j+1) \left( 
\frac{v \bar{u}}{r^2} \right)^{j}= \frac{1}{\pi r^2}
\frac{1-a^N (1+ N-a N)}{(1-a)^2}
\, ,  \quad  \mbox{in $D_r$}\ , 
\nn \\
\ka_{N}(v , \bar{u})&& =\frac{1}{2 \pi r}\sum_{j=0}^{N-1} \left( 
\frac{v \bar{u}}{r^2} \right)^{j}= \frac{1}{2 \pi r}\frac{1-a^N}{(1-a)}
\, , \quad  \mbox{in $C_r$ \, ,} \quad a=\frac{v\bar{u}}{r^2} \, .
\eea
Actually, all domains and weight functions in the complex plane which
are invariant under the rotation $z \to e^{i \theta} z, \forall
\theta\in [0,2 \pi]$ (i.e. they are function only of the radial
coordinate $\rho=|z|$), give orthogonal polynomials which are monic
powers, i.e. $\tilde{P}_j(z)=z^j$ with normalization coefficients
given by
\be
h_j=2 \pi \int_0^{r} d\rho\,  \rho^{2j+1} \, \ w(\rho,\rho)\, .
\ee
The Gaussian case $w_G(z,\bar{z})$ we considered above belongs to this
class, and therefore the same arguments apply here.  In particular,
eq. (\ref{LGin}) holds exactly for all the elements of this class,
since it is independent of the particular expression for $h_j$.

A more interesting case is for a bounded domain $D$ which is not
rotationally invariant, e.g. the ellipse
\be
D^{ell}_{r}=\{z \in \mathbb{C}: |z|< \frac{1}{2}(r e^{i \theta}+ c^2 r^{-1}
 e^{-i \theta} ) ,\  \forall \theta \in [0,2 \pi] \} \, ,  
\quad r \geq c \geq 0 \, 
\ee
or its boundary $C^{ell}_r=\partial D^{ell}_r$. The shape of the
ellipse is parameterized by two parameters, $r$ and the semi-focal
length $c$. Obviously the semi-minor/major axis lengths are equal to
$r-c^2 r^{-1},r+c^2 r^{-1}$, respectively. It is easy to see that in
this case the monic orthogonal polynomials with respect to a flat
measure $w(z,\bar{z})=1$, are given by the Chebyshev polynomials of
the second kind:
\be
\tilde{P}_j(z)=\left(\frac{c}{2} \right)^j U_{j} \left( \frac{z}{c} \right) 
\, ,  
\quad h_j=\frac{\pi}{2^{2j+2}(j+1)}  \left( r^{2j+2}- \left( \frac {c^2} 
{r} \right)^{2j+2}  \right)  \, .
\label{Ellipse}
\ee
It is easy to check that eq. (\ref{Ellipse}) reduces to eq. (\ref{Disk}) 
in the limit $c \to 0$. The kernel reads:
\be
\ka_N(v,\bar{u})= 
\frac{4}{\pi}\sum_{j=0}^{N-1} 
\frac{c^{2j}(j+1)}{r^{2j+2}-(c^2 r^{-1})^{2 j+2}}
 U_{j} \left( \frac{v}{c} \right) 
U_{j} \left( \frac{\bar{u}}{c} \right)  \ .
\ee

The final example we want to give are the generalized Laguerre
polynomials in the complex plane as recently introduced in
\cite{A02}. For this chiral complex eigenvalue model no matrix
representation is known so far and we have to slightly modify our
definitions (\ref{Zev}) and (\ref{OPdef}).  The partition function is
given by
\be
{\cal Z}_N^{ch} \ \equiv\ \int_D\prod_{i=1}^N \left( d^2z_i\ 
w_{ch}(z_i,\bar{z}_i)\right) \left|\Delta_N(z^2)\right|^2\  \, ,
\label{Zchev}
\ee
with $D$ being the full complex plane and a weight 
\be
w_{ch}(z,\bar{z})\ \equiv\ |z|^{2a+1}
\exp\left[-\frac{1}{1-\tau^2} \left( |z|^2
\ -\ \frac{\tau}{2}(z^2+\bar{z}^2)\right)\right] \ , \ \ 
\ a>-\frac12 \in \mathbb{R}\ .
\label{wch}
\ee
We could also allow for more general weight functions and domain $D$,
the crucial point for the model to be chiral being the squared
argument inside the Vandermonde determinant. The orthogonal
polynomials are then necessarily polynomials in $z^2$:
\be
h_k\delta_{kl}\ = \ \int_D d^2z \ w(z,\bar{z})\tilde{P}^{ch}_k(z^2)
\overline{\tilde{P}^{ch}_l(z^2)} \ .
\label{chOPdef}
\ee
The characteristic polynomials consequently have to be defined in
terms of the eigenvalues
\be
\left\langle \prod_{i=1}^N\left(\prod_{j=1}^K (v_j^2-z_i^2)\ 
\prod_{l=1}^L (\bar{u}_l^2-\bar{z}_i^2)
\right)\right\rangle_N \, .
\ee
The whole proof of Theorem (\ref{Th}) goes through as before,
replacing all arguments by their square on the right hand side.

The finite-$N$ orthogonal polynomials for our example eq. (\ref{wch})
are given by the generalized Laguerre polynomials \cite{A02II}
\be
\tilde{P}^{ch}_k(z^2)\ =\ (-1)^k\frac{\Gamma(a+1)k!}{\Gamma(a+k+1)} 
\ L_k^a\left(\frac{z^2}{2\tau}\right)\ .
\label{Lague}
\ee
Their norm can be calculated to be 
\bea
h_k &=& \frac{f^a(\tau)\Gamma(a+1)k!}{\tau^{2k}\Gamma(a+k+1)}\ , \nn\\
f^a(\tau)&=& \int_Dd^2z\ w_{ch}(z,\bar{z})\ =\ \pi\ 
\Gamma\left(a+\frac32\right)
(1-\tau^2)^{\frac{a}{2}+\frac34}\ {\cal P}_{a+\frac12}
\left(\frac{1}{\sqrt{1-\tau^2}}\right) \ ,
\label{chnorm}
\eea
where ${\cal P}_{a+\frac12}(x)$ is the Legendre function.
The bare kernel is thus reading 
\be
\ka_N(z,\bar{u})\ =\
\frac{1}{f^a(\tau)}
\sum_{k=0}^{N-1} \frac{\Gamma(a+1)k!}{\Gamma(a+k+1)}\ \tau^{2k}
 L_k^a\left(\frac{z^2}{2\tau}\right) 
L_k^a\left(\frac{\bar{u}^2}{2\tau}\right) \, .
\label{kerLague}
\ee

\subsection{A duality relation}

In this subsection we wish to prove a duality relation that relates
products of characteristic polynomials of different (finite) matrix
size.  This generalizes the relation found in \cite{MN,FW} to complex
matrices.  The proof goes back to a relation between determinants of
Hermite polynomials \cite{MN} which we will use. Therefore our duality
relation will only hold for the weight eq. (\ref{wHerm}) having
Hermite polynomials in the complex plane.

To begin with we consider Theorem eq. (\ref{Th}) at $L=0$ where only
polynomials and no kernels appear. Next we take the limit of equal
arguments $v_i=v$, $\forall i=1,\ldots,K$. We obtain
\be
\left\langle \, \det[v-J]^K \,
\right\rangle_N
\ =\  \frac{1}{\prod_{j=1}^{K-1}  j!}
\det_{1\leq l,m\leq K}[\ \partial^{(l-1)}_v \tilde{P}_{N+m-1}(v) \ ] \ .
\label{deg}
\ee
This looks exactly as for the hermitian matrix model (e.g. in 
\cite{MN}).
Next we insert the orthonormalized Hermite polynomials eq. (\ref{Herm}) 
for the chosen weight eq. (\ref{wHerm}) and obtain
\bea
\left\langle \, \det[v-J]^K \,
\right\rangle_N
&=&  \frac{1}{\prod_{j=1}^{K-1}j!}
\left(\frac{\sqrt{2\tau}}{2}\right)^{NK+K(K-1)/2}
\sqrt{2\tau}^{-K(K-1)/2} 
\det_{1\leq l,m\leq K}
\left[  H_{N+m-1}^{(l-1)}\left(\frac{v}{\sqrt{2\tau}}\right) 
\right] \nn\\
&=&  \frac{1}{\prod_{j=1}^{K-1}j!}
\left(\frac{\sqrt{2\tau}}{2}\right)^{NK}
\left(\frac{-1}{2}\right)^{K(K-1)/2}
\det_{1\leq l,m\leq K}
\left[ H_{N+m+l-2}\left(\frac{v}{\sqrt{2\tau}}\right) \right]. 
\label{deg2}
\eea
In the first step we have taken out all factors of the determinant
stemming from the monic normalization and the differentiation, where
$H_{N+m-1}^{(l-1)}(v/\sqrt{2\tau})$ denotes the derivative with
respect to the full argument $v/\sqrt{2\tau}$. In the second step we
have used the following relation for the Hermite polynomials,
\be
H'_N(x) \ =\ 2xH_N(x) - H_{N+1}(x) \ ,
\ee
which holds algebraically for any $x$ being real or complex. We have
thus shifted the derivative into the index of the polynomials, taking
out all signs.  We can now use the following relation among
determinants of Hermite polynomials as being shown in
\cite{MN}\footnote{In \cite{MN} this relation was shown for the
determinant of $C_n(x)\equiv 2^{-n}H_n(x)$, but the factors of 2 can
be easily taken out.  We further note a typo in eq. (3.16) there.}
\be
\frac{\det_{1\leq l,m\leq K}[\ H_{N+m+l-2}(x)^{}\ ]
}{(-2)^{K(K-1)/2}\prod_{j=1}^{K-1}j!}
\ =\ (-i)^{KN}
\frac{\det_{1\leq l,m\leq N}[\ H_{K+m+l-2}(ix)^{} \ ]
}{(-2)^{N(N-1)/2}\prod_{j=1}^{N-1}j!} \ .
\label{Hid}
\ee
This relation was proved in \cite{MN} for real $x$ only. However, the
relation is purely algebraic as it relates two polynomials after
multiplying out the two determinants. We can therefore also allow for
$x$ to be complex, in particular $x=v/\sqrt{2\tau}$. Note also the
factor of $i$ in the argument on the right hand side.

Replacing the determinant of size $K\times K$ 
in eq. (\ref{deg2}) by a determinant of size $N\times N$ with the help of eq.
(\ref{Hid}), we arrive at the following remarkable identity:
\be
\left\langle \, \det[\,v-J\,]^K \,\right\rangle_N 
\ =\  (-i)^{KN}\left\langle \, \det[\, iv-J\,]^N \,\right\rangle_K \ ,
\label{dual}
\ee
interchanging the power of the product with the matrix size. 
It is identical to the corresponding equation in \cite{MN} replacing the 
complex matrix $J$ and argument $v$ by the respective hermitian matrix $H$
and real argument $x$. 
The normalization in eq. (\ref{dual}) can be easily checked taking the limit
$v\to\infty$. Another peculiar case is 
$N=K$ at $v=0$. Still, the duality relation is satisfied as both expectation 
values which are now equal vanish for odd $N=K$ due to symmetry.

The relation eq. (\ref{dual}) trivially extends to the Ginibre ensemble 
eq. (\ref{wGin}), which can be obtained by taking $\tau\to0$ on 
the Hermite polynomials eq. (\ref{Herm}). It is because in the Ginibre case 
all expectation values of powers of $J$ vanish. The same argument applies 
for all weights with spherical symmetry and monomials as orthogonal 
polynomials, as in eqs. (\ref{Disk}) and (\ref{Circle}). 
We suspect that the duality also holds for other weight functions.

\sect{The microscopic large-$N$ limit and universality}\label{largeN}

The explicit examples given in the last Section are a good starting
point for the analysis of the asymptotic large-$N$ limit. To this aim
we first rescale the complex eigenvalues by $z_i\to\sqrt{N}\,z_i$ in
order to have $N$ explicitly in the exponent of the weight functions
eqs. (\ref{wHerm}) or (\ref{wch}), to which we will restrict in this
Section.  Then we focus on the microscopic large-$N$ limit where the
correlations of eigenvalues at small distance O$(|z_i-z_j|)\ll 1$ are
considered. For spectral correlators at a distance of O(1) (the
so-called macroscopic limit) we refer to \cite{WZ,EK}.

Furthermore, one has to distinguish two different ways of taking the
microscopic $N\to\infty$ limit \cite{FKS}. Starting from a general
weight such as in eq. (\ref{expweight}) or simply from the Ginibre
ensemble eq. (\ref{wGin}) we will always end up in a regime which is
called {\it strong non-hermiticity}. Introducing an additional
parameter $\tau\in[0,1]$ as in eq. (\ref{wHerm}) will allow us to
reach also the regime of {\it weak non-hermiticity}, when taking
simultaneously $N\to\infty$ and $(1-\tau)\to0$ while keeping the
product fixed. The latter limit introduced in \cite{FKS} permits to
smoothly interpolate between correlations of real eigenvalues on one
hand and complex eigenvalues in the strong non-hermiticity limit on
the other hand.

Let us begin with the weak non-hermiticity limit. Here we focus on a
weight function of the form of eq. (\ref{expweight}), which is in
terms of eigenvalues
\be
w(z,\bar{z})\ \equiv\ \exp\left[-N\left(
\frac{1}{1-\tau^2}\left(|z|^2-\frac{\tau}{2}(z^2
  +\bar{z}^2)\right)
\ +\ \frac12\sum_{k=2}^d \frac{g_{2k}}{2k} (z^{2k}+ \bar{z}^{2k}) 
\right)\right] \, ,
\label{weakweight}
\ee
which generalizes the weight eq. (\ref{wHerm})\footnote{Strictly
speaking the integral over the weight function eq. (\ref{weakweight})
cannot be made convergent for any choice of signs of the coupling
constants $g_{2k}$.  Consequently one would either have to restrict
the domain $D$ to be compact or to introduce a small term
$\sim\epsilon |z|^{2d+2}$ to achieve convergence.  However, in the
weakly non-hermitian large-$N$ limit the support of the eigenvalues
does not only become compact but also gets projected onto the real
line. The correlators depend on complex variables only
microscopically.  For that reason we ignore the aforementioned
technical subtleties.}.  Here the $g_{2k}\in \mathbb{R}$ are real
coupling constants and $\tau\in\ [0,1]$ measures the degree of
non-hermiticity as explained after eq. (\ref{wHerm}).  The weak
non-hermiticity limit is defined by taking \cite{FKS}
\be
\lim_{\stackrel{N\to\infty}{\tau\to1}}N(1-\tau^2) \equiv \al^2 
\label{weaklim}
\ee
to be constant. At the same time we rescale the complex eigenvalues according 
to 
\be
Nz\ =\ N(\re z +i\im z) \ \equiv\  \xi \ \ ,
\label{microlim}
\ee
which is the microscopic limit at the origin since $z\to0$ while
$N\to\infty$.  
It is important to note that the scaling is done with the same
power of $N$ as for real eigenvalues, in contrast to the strong
non-hermiticity limit (see eq.  (\ref{strongmicrolim}) below).  This
has important consequences on the asymptotic limit of the orthogonal
polynomials as well.

We will now give the asymptotic results for the orthonormal
polynomials of eq. (\ref{weakweight}) and their kernel as they were
derived in
\cite{A02II}:
\bea
\lim_{\stackrel{n,N\to\infty}{\tau\to1}} 
P_n\left(z=\frac{\xi}{N}\right)
=\left[\sqrt{2}\,Nu(t)'\al^{-1}\pi^{-\frac32}
\right]^{\frac12} \mbox{e}^{-\frac{\al^2}{4}u(t)^2}
\left\{
\begin{array}{ll}
\cos(u(t)\xi) & n\ \mbox{even}\\
\sin(u(t)\xi) & n\ \mbox{odd}\\
\end{array}
\right. ,\ t=\frac{n}{N}\ .
\label{OPfinal}
\eea
In \cite{A02II} the full kernel eq. (\ref{Kerdef}) including the 
weights was given. Using the fact that the asymptotic limit of the 
weight function reads 
\be
w(z,\bar{z})\ \to\ \exp\left[-\frac{2}{\al^2}(\im z)^2\right] \, ,
\ee
we obtain for $\ka_N(z_1,\bar{z_2})$ appearing in Theorem (\ref{Th})
\be
\lim_{N\to\infty} \frac{1}{N^2}\ka_N\left(z_1=\frac{\xi_1}{N},\bar{z_2}=
\frac{\bar{\xi}_2}{N}\right)\ =\ 
\frac{2}{\al\pi}\int_0^{\pi\rho_1(0)} \frac{du}{\sqrt{2\pi}}\ 
\mbox{e}^{-\frac{\al^2}{2}u^2}
\cos\left[u(\xi_1-\bar{\xi}_2)\right] \ .
\label{kerfinal}
\ee
Here we have introduced the spectral density of real eigenvalues $\rho_t(x)$ 
corresponding to the hermitian limit of the potential in eq. 
(\ref{weakweight}):
\bea
\rho_t(x) &\equiv&
\frac{1}{2\pi} \sum_{j=1}^m \frac{g_{2j}}{t} \sum_{k=0}^{j-1} 
{{2k}\choose{k}}
r(t)^{k} x^{2(j-k-1)}
\sqrt{4r(t)-x^2}  \ ,
\label{rho}\\
1&=&\frac12 \sum_{k=1}^m \frac{g_{2k}}{t} {{2k}\choose{k}}r(t)^{k} \, .
\nn
\eea
The function $u(t)$ in eq. (\ref{OPfinal}) is related to the spectral density
 by 
\be
\pi t\rho_t(0) \ \equiv\ u(t) \ ,
\label{urho}
\ee
Finally the asymptotic of the norm of the polynomials is:
\be
h_N\  \to\ \sqrt{\frac{u'(t)}{\al}}\ .
\label{Nnorm}
\ee
The remaining $N$-dependence of the norm $h_N\sim N!$ in the
prefactors of eq. (\ref{Th}) has to be canceled by a suitable
normalization. Inside the determinant it is needed when passing from
the monic to the orthonormal polynomials to make the asymptotic limit
of the latter well defined.  To summarize we have determined the weakly
non-hermitian large-$N$ limit of the characteristic polynomials as
stated in Theorem eq. (\ref{Th}), when inserting eqs. (\ref{OPfinal}),
(\ref{kerfinal}) and (\ref{Nnorm}).  Hence it is also universal.

In \cite{FKS} the weak non-hermiticity limit away from the origin was
also analyzed, for the Gaussian weight of eq. (\ref{wHerm}).  The
corresponding kernel is exactly of the same form of
eq. (\ref{kerfinal}), with replacing $\pi\rho_1(0)\to\sqrt{1-X^2/4}$
which is the Wigner semicircle density at the point of investigation
$X\in\ \mathbb{R}$.  While the norms $h_N$ still behave as in
eq. (\ref{Nnorm}) with $u(t)=\sqrt{2t}$ the asymptotic polynomials
will acquire an additional oscillating phase
$\sim\cos(\sqrt{2t}(\xi+NX))$ as we now microscopically rescale the
deviation from point $X$: $z=X+\xi/N$, with $z,\xi\in\ \mathbb{C}$.
In other words their asymptotic limit does no longer exist.  Therefore
only for $K=L$ a smooth limit is found at weak non-hermiticity away
from the origin (a similar situation is found below at strong
non-hermiticity). Furthermore the question of universality away from
the origin remains open.

We now turn to the microscopic limit at strong non-hermiticity where we keep 
$\tau\in\ [0,1)$  fixed while $N\to\infty$.
The microscopic origin scaling limit is defined here by letting $z\to0$ and 
$N\to\infty$ keeping 
\be
\sqrt{N}\ z\ =\ \sqrt{N}(\re z +i\im z) \ \equiv\  \xi \ \ ,
\label{strongmicrolim}
\ee
constant.  The reason for the different scaling is that now the
eigenvalues will become dense on a compact domain which truly extends
into the complex plane. This has an immediate consequence as the
asymptotic large-$N$ of the orthogonal polynomials no longer
exists. This can be seen in the simplest example of the Ginibre
ensemble eq. (\ref{wGin}) or of the bounded domains eq. (\ref{Disk}),
(\ref{Circle}). The polynomials $P_N(z)=z^N/N!$ from eq. (\ref{Gin})
will certainly no longer have a smooth large-$N$ expansion. The same
problem occurs for the Hermite polynomials in the complex
plane. Expanding for example the even powers,
$H_{2j}(z\sqrt{\frac{N}{2\tau}})\sim\cos[\sqrt{4j+1}\
z\sqrt{\frac{N}{2\tau}}]$ it can be seen that while the weak scaling
limit eq.  (\ref{microlim}) is smooth (keeping $j/N=t$ fixed) the
strong limit from eq.  (\ref{strongmicrolim}) does not exist.  While
on the real line the existence of an asymptotic expansion of the
polynomials and the kernel are directly related through the
Christoffel-Darboux formula eq. (\ref{CD}) this is no longer the case
here, since such a relation does not hold in general in the complex
plane.  However the asymptotic limit of the kernel eq. (\ref{kerGin})
still exist and it is given by the exponential function being smooth.
The result for the weight in eq. (\ref{wHerm}) reads after splitting
of the weight factors \cite{FKS}:
\be
\lim_{N\to\infty} \frac{1}{N}
\ka_N\left(z_1=\frac{\xi_1}{\sqrt{N}},\bar{z}_2=\frac{\bar{\xi}_2}{\sqrt{N}}
\right)\ =\ 
\frac{1}{\pi(1-\tau^2)} \exp\left[\frac{1}{(1-\tau^2)}
\left(\xi_1\bar{\xi}_2 -\frac{\tau}{2}(\xi_1^2+\bar{\xi}_2^2)\right)\right]\ .
\label{strongkerfinal}
\ee
For $\tau\to0$ we recover the exponential Ginibre kernel \cite{Gin}.
Taking into account the rescaled norms $h_N\sim\sqrt{1-\tau^2}$ we can
read off the asymptotic strongly non-hermitian large-$N$ limit of our
characteristic polynomials eq. (\ref{Th}) for $K=L$.  The question of
universality of eq. (\ref{strongkerfinal}) for more general weight
functions remains an open question.  This has to be compared with the
situation in the macroscopic large-$N$ limit. There, it has been
recently shown that the two-point resolvent is universal but the
three- and four-point function are not \cite{WZ}.

The same analysis for the microscopic large-$N$ limit at weak and
strong non-hermiticity can be done for the chiral ensemble of complex
eigenvalues introduced in eqs. (\ref{Zchev}) and (\ref{wch}). The
asymptotic expressions for the norms, polynomials and kernel in the
weak limit, and the norms and kernel in the strong limit have already
been explicitly given in \cite{A02II} and we will not repeat them
here.  We find the same phenomenon as above that only the case $K=L$
at strong non-hermiticity makes sense asymptotically.  At weak
non-hermiticity there are no such restrictions. The issue of
universality for the chiral ensemble remains an open problem.

We finish this Section with a remark: in the mathematical literature
the large-$N$ macroscopic limit of our kernel $K_N(v,\bar{u})$ is
known as Bergman kernel function or Szeg{\"o} kernel function,
according to the case where the weight function is defined on a domain
in the complex plane or on its boundary. Both these kernels are
objects of fundamental importance in the theory of conformal mapping
\cite{stahl}.


\sect{The BMN large-$N$ limit}\label{BMNlim}

In this Section we apply the results of Sections
\ref{Appl} and \ref{finiteN} to the study of objects 
like  
$\langle \tr J^K\tr J^{\dagger\, K}\rangle_N$ in the limit where both 
$N$ and $K$ become large. Such  
quantities 
have become of interest in the 
correspondence between pp-wave strings 
and supersymmetric Yang-Mills theory 
\cite{plefki}. 
In this case the expectation values are evaluated with respect to the
weight of the Ginibre ensemble eq. (\ref{wGin}).  It has been found
that besides the usual (macroscopic) large-$N$ limit, one can consider
also $K,N\to\infty$ such that $\frac{K^2}{N}$ is finite. This limit is
called BMN limit after \cite{BMN}.  In a sense in this limit the
expectation value becomes part of the matrix model action and the
weight is no longer truly Gaussian as in eq. (\ref{wGin}). It would be
very interesting to study more general insertions to the matrix model
action and to see if such models permit a large-$N$ BMN limit.  In
this Section we will study the weight eq. (\ref{wHerm}) for the
Hermite polynomials as an example.

It is convenient to introduce 
products of resolvent operators such as
\be
W(z,\bar{u})\ \equiv\ \left\langle 
\tr \frac{1}{z-J}\ \tr\frac{1}{\bar{u}-\Jd}
\right\rangle\ =\ \frac{1}{z\bar{u}}\sum_{l,m=0}^\infty
\frac{1}{z^l \ \bar{u}^m }
\left\langle \tr J^l \
\tr J^{\dagger\, m}
\right\rangle\ ,
\label{resolvent}
\ee
in order to generate expectation values of product of traces of powers
of matrices. However, at finite-$N$ only few closed formulas for a
small number of traces are known so far, such as
\cite{plefki} 
\be
\left\langle \tr J^l
\tr J^{\dagger\, m}
\right\rangle\ =\ \delta_{lm} 
\left(\frac{\Gamma(N+l+1)}{\Gamma(N)}\ -\ 
\frac{\Gamma(N+1)}{\Gamma(N-l)}\right)  \, , 
\label{2pt} 
\ee
for the Ginibre weight eq. (\ref{wGin}). 
More general multi-point resolvents 
are known only 
in the asymptotic large-$N$ limit so far \cite{EK}.

The products of characteristic polynomials we have calculated in eq. (\ref{Th})
 are generating functions that may be used to extract 
information about the spectrum.
In fact they generate all the elementary symmetric functions $c_r$ 
of the eigenvalues:
\bea
\det[v+J] &=&  v^N \sum_{r=0}^N c_r v^{-r}\ ,\nn\\
c_r       &\equiv& \sum_{i_1<\ldots<i_r}^N z_{i_1}\cdots z_{i_r}\ ,\ \ \ 
r=1,\ldots,N\ \ .
\label{cr}
\eea
Similarly we have $\det[\bar{u}+\Jd]= \bar{u}^N \sum_{r=0}^N \bar{c}_r
\bar{u}^{-r}$ for the hermitian conjugate.
The elementary symmetric functions $c_r$ can be expressed 
by the power sums 
\be
p_r\ \equiv\ \sum_{i=1}^N z_i^r\ =\ \tr J^r\ ,\ \  r=1,\ldots,N\ \ .
\label{pr}
\ee
Namely, the two sets are related by
\be
c_r = \frac{1}{r!} \det_{1\leq i,j,\leq r}[S_{ij}] \, , \quad 
\label{cprel} 
S_{ij} = \left\{\begin{array}{cl}
p_{j-i+1} & \mbox{if}\ j\geq i\\
j         & \mbox{if}\ j= i-1\\
0         & \mbox{if}\ j< i\\
\end{array}\right.\ .
\ee

As a first example we can apply the Theorem eq. (\ref{Th}) for $K=L$
using the finite-$N$ result eq. (\ref{kerGin}) for the Gaussian 
weight:
\bea
\langle\det[v-J]\det[\bar{u}-\Jd]\rangle_N &=& h_N \ka_{N+1}(v,\bar{u})\ =\ 
N!\sum_{j=0}^N \frac{1}{j!}(v\bar{u})^j \nn\\
&=&\sum_{l,m=0}^N
\langle c_l\, \bar{c}_m\rangle_N \  (-v)^{N-l} (-\bar{u})^{N-m} \ . 
\label{ccbardef}
\eea
Comparing coefficients we obtain 
\be
\langle  c_l\, \bar{c}_m\rangle_N \ =\ \delta_{lm} \frac{N!}{(N-l)!} \ .
\label{ccbar}
\ee
In the large-$N$ limit it yields
\bea
\langle  c_l\, \bar{c}_l\rangle_N 
&=& N^l \left( 1-\frac{l(l-1)}{2N}+\frac{1}{12}
{{l}\choose{2}}\left(3(l-1)^2-l-1\right)\frac{1}{N^2}+\ldots\right) \nn\\
&=& N^l \left(1 + \sum_{h=1}^\infty a_h\frac{l^{2h}}{N^h}+\ldots\right) \ ,
\label{ccexpand}
\eea
and consequently an expansion in $l^2/N$ exists. We note that the
expansion goes in powers of $\frac1N$ instead of the usual genus
expansion in powers of $\frac{1}{N^2}$. This is due to the fact that
eq. (\ref{ccbar}) contains expectation values of all combinations
$\langle p_{i_1}\cdots p_{i_r}\bar{p}_{j_1}\cdots\bar{p}_{j_r}\rangle_N$
with $\sum_{s=1}^ri_s=\sum_{s=1}^rj_s=r$.  So far such expectation
values were not known explicitly for finite-$N$ in a closed form.  If
we could manage to disentangle all expectation values of $p_r$ and
$\bar{p}_r$ alone, the individual terms would all have a genus
expansion in $\frac{1}{N^2}$, with different prefactors in $N$
however.  The fact that the BMN limit of eq. (\ref{ccbar}) exists is
quite remarkable.  This holds for the Ginibre ensemble. We wish to
mention however, that for a more general weight function the expansion
is generically in powers of $\frac1N$. For example the next to leading
order of the free energy in $\frac1N$ can be calculated \cite{WZ}.

The difficulty to extract formulas like eq. (\ref{2pt}) from
eq. (\ref{ccbar}) is that each elementary symmetric functions $c_r$
depends on all lower powers $p_1,\ldots,p_r$ through the determinant
in eq. (\ref{cprel}).  The fact that the $c_r$ can be constructed
recursively, $nc_n=\sum_{r=1}^n(-1)^{r-1}p_r c_{n-r}$, is also not
enough to extract each single term. We have therefore not been able to
derive new finite-$N$ expectation values of traces alone, that would
generalize eq. (\ref{2pt}). However, we have explicitly checked in few
cases that eq. (\ref{ccbar}) leads to eq. (\ref{2pt}).

We will now address the question whether the existence of the BMN
limit is a generic property of complex matrix models or if it is
specific to the Ginibre ensemble. Therefore we look at the ensemble
with weight eq. (\ref{wHerm}) of Hermite polynomials. In this model
single powers of traces $\langle\tr J^l\rangle$ do not vanish any
longer for $\tau\neq0$. The simplest nontrivial characteristic
polynomial is given by
\be
\langle\det[v-J]\rangle_N\ =\ \tilde{P}_N(v)\ =\ 
\left(\frac{\tau}{2}\right)^{\frac{N}{2}} 
H_N\left(\frac{v}{\sqrt{2\tau}}\right) \ ,
\label{chHerm}
\ee
where we have used eq. (\ref{Herm}).  In order to read off the
coefficients from eq. (\ref{cr}) by expanding the right hand side we 
distinguish between even and odd $N$. For $N=2n$ we obtain
\bea
\langle\det[v-J]\rangle_{2n} &=&  \sum_{l=0}^n 
\langle\ c_{2n-2l} \rangle_{2n}  \ (-v)^{2l} \nn\\
&=& \sum_{l=0}^n (-1)^{n+l} (2\tau)^{n-l} 
{{n}\choose{l}}
\frac{\Gamma(n+\frac12)}{\Gamma(l+\frac12)}\ 
v^{2l}\ ,
\label{chHermeven}
\eea
where we have used the fact that our weight is even. 
Repeating the same analysis for $N=2n+1$ and using an identity for 
the Gamma function we arrive at 
\be
\langle\, c_{2l}\, \rangle_N \ =\ \left(-\frac{\tau}{2}\right)^l
\frac{N!}{l!(N-2l)!} \, , \quad \langle\, c_{2l+1}\, \rangle_N \ =0 \, , 
\label{ccHermexp} 
\ee
which holds for any $N$. Looking at eq. (\ref{ccbar}) we can see that
again the limit $l,N\to\infty$ with $l^2/N$ fixed exists. Such a
large-$N$ limit thus seems to be a generic property of complex matrix
models.  We note however, that the expectation value
eq. (\ref{ccHermexp}) has to be properly normalized since both
$\tau^l$ and $1/l!$ vanish for $0<\tau<1$ at large-$l$.


\sect{Conclusions}

We have studied products of characteristic polynomials of complex
matrices and their hermitian conjugate. For hermitian matrix models
several equivalent determinant formulas hold for characteristic
polynomials.  They contain orthogonal polynomials and their kernel,
which are related through the Christoffel-Darboux formula.  For
complex matrices this is no longer the case. The number of
characteristic polynomials and their conjugate completely fixes the
result which in general is a determinant of both the orthogonal
polynomials in the complex plane and their respective kernel.  If we
only consider characteristic polynomials of complex matrices without
their conjugate our result reduces to a determinant over polynomials
only, which is identical to the result of the hermitian matrix model.
This does not come as a surprise since our proof only uses the
existence of orthogonal polynomials in the complex plane and holds
despite of the breakdown of the three-step recursion relation and the
Christoffel-Darboux identity.

Next we gave an interpretation of our results in terms of orthogonal
polynomials in the complex plane, with weights including determinants,
and their corresponding kernels.  We have provided explicit examples
for complex polynomials at finite-$N$ such as Hermite, Laguerre or
Chebyshev polynomials in the complex plane.  From them we have
constructed orthogonal polynomials for a class of more general
weights.

At finite-$N$ we proved a duality relation for the weight with Hermite
polynomials in the complex plane. It relates expectation values of
different powers of characteristic polynomials with respect to a
different matrix size, generalizing a known duality for the Gaussian
Hermitian matrix model.

We have also studied the microscopic large-$N$ limit of characteristic
polynomials when the arguments are rescaled with the matrix size $N$.
In the weak non-hermiticity limit all products of characteristic
polynomials are universal for a large class of weight functions.  The
strongly non-hermitian large-$N$ limit is smooth only for products
with an equal number of characteristic polynomials and their hermitian
conjugate.  This is again a peculiar implication of the breakdown of
the Christoffel-Darboux identity. Although in general the orthogonal
polynomials do no longer have a smooth asymptotic limit, their kernel
still can and does have a smooth limit.  The issue of universality of
the large-$N$ kernel at strong non-hermiticity remains open at
present.

Furthermore we have argued that the existence of the BMN large-$N$
limit is a generic phenomenon for complex matrix models.  In
using the fact that characteristic polynomials generate the elementary
symmetric functions we have shown that for quite general expectation
values such a limit exists. In principle from our finite-$N$ results
one can compute iteratively expectation values of single powers of
traces alone. However it would be more useful to find a closed expression for that, if any. 

It would be also very interesting to extend our analysis to include
negative powers of moments in the expectation values. We suspect that
a result similar to that of hermitian matrices holds, where the Cauchy
transformation of the orthogonal polynomials occurs.  This would
provide us with an alternative way to evaluate complex eigenvalue
correlations at finite-$N$ and infinite-$N$.

\indent

\noindent
\underline{Acknowledgments}: We wish to thank A. D'Adda, F. David and 
M.L. Mehta for useful discussions and P. Forrester for correspondence.
This work was supported by the European network on ``Discrete Random
Geometries'' HPRN-CT-1999-00161 EUROGRID and by a Heisenberg
fellowship of the Deutsche Forschungsgemeinschaft.


\indent

\end{document}